\documentclass[aps,prl,twocolumn,superscriptaddress]{revtex4}
\usepackage{graphicx}

\newcommand{\be}{\begin{equation}}
\newcommand{\ee}{\end{equation}}
\newcommand{\ba}{\begin{eqnarray}}
\newcommand{\ea}{\end{eqnarray}}
\newcommand{\baa}{\begin{eqnarray*}}
\newcommand{\eaa}{\end{eqnarray*}}

\def\be{\begin{equation}}
\def\ee{\end{equation}}
\def\bea{\begin{eqnarray}}
\def\eea{\end{eqnarray}}

\def\C60{A$_x$C$_{60}$}

\def\HgCu3{HgCa$_2$Cu$_3$O$_{8+y}$}
\def\HgCu4{HgBa$_2$Ca$_3$Cu$_4$O$_{10+y}$}
\def\TlCu{Tl$_2$Ba$_2$CuO$_{6+\delta}$}
\def\TlCu3{Tl$_2$Ba$_2$Ca$_2$Cu$_3$O$_{10+y}$}
\def\TlCu4{Tl$_2$Ba$_2$Ca$_3$Cu$_4$O$_{12+y}$}

\def\BiCu3{Bi$_2$Sr$_2$Ca$_{2}$Cu$_3$O$_y$}

\def\8LSCO{La$_{1.88}$Sr$_{.12}$CuO$_4$}
\def\110LNSCO{La$_{1.5}$Nd$_{0.4}$Sr$_{0.1}$CuO$_{4}$}
\def\stage4LCO{La$_{2}$CuO$_{4+\delta}$}
\def\Y248{YBa$_2$Cu$_4$O$_8$}

\def\NbSe2{NbSe$_2$}
\def\TaSe2{TaSe$_2$}
\def\TiSe2{TiSe$_2$}

\begin{document}
\title{   Mechanism for  Odd Parity Superconductivity in Iron-Based Superconductors }
\author{Jiangping Hu}
\email{jphu@iphy.ac.cn/hu4@purdue.edu}
\affiliation{Beijing National
Laboratory for Condensed Matter Physics, Institute of Physics,
Chinese Academy of Sciences, Beijing 100080,
China}
\affiliation{Department of Physics, Purdue University, West
Lafayette, Indiana 47907, USA}
 \author{Ningning Hao}
\affiliation{Beijing National
Laboratory for Condensed Matter Physics, Institute of Physics,
Chinese Academy of Sciences, Beijing 100080,
China}
\affiliation{Department of Physics, Purdue University, West
Lafayette, Indiana 47907, USA}
\author{Xianxin Wu}
\affiliation{Beijing National
Laboratory for Condensed Matter Physics, Institute of Physics,
Chinese Academy of Sciences, Beijing 100080,
China}
\begin{abstract} Under the assumption that superconducting pairing is driven by local d-p hybridization, we show that the superconducting state in iron-based superconductors is classified as an odd parity s-wave spin-singlet pairing state in a single trilayer FeAs/Se, the building block of the materials. In a low energy effective model with only  d-orbitals in an iron square bipartite lattice,  the superconducting order parameter in this state is a combination of a s-wave normal pairing between two sublattices and  a s-wave $\eta$-pairing within the sublattices. Parity conservation was violated in proposed superconducting states in the past.  The results demonstrate  iron-based superconductors being a new quantum state of matter and suggest that  a measurement of odd parity can establish fundamental principles related to high temperature superconducting mechanism. 
   \end{abstract}
\maketitle
 In  a strongly correlated electron system, major physics is   determined locally in real space. Important properties, such as pairing symmetry in a superconducting state, are expected to be robust against small variation of Fermi surfaces in reciprocal space. Although superconducting mechanism related to high temperature superconductors (high $T_c$)  is still unsettled, the robust d-wave pairing symmetry in cuprates\cite{Tsuei2000} 
can be understood under this principle. 

Is this principle still held for iron-based superconductors\cite{Hosono,ChenXH, wangnl2008,ironbook}? Namely, do all iron-based superconductors possess one universal pairing state?  Unlike cuprates, the answer to this question is highly controversial because different theoretical approaches have provided different answers and no universal state has been identified\cite{review-hirschfeld2011}.  Nevertheless,  as local electronic structures in all families of iron-based superconductors  are  almost identical and phase diagrams  are smooth against doping\cite{Johnstonreview,review-hirschfeld2011,ironbook}, it is hard to argue that the materials can approach many different superconducting ground states.
% In  many studies based on weak coupling approaches,  pairing symmetries in iron-based superconductors  become very sensitive to doping and Fermi surface topologies.  In contrast,  s-wave is proposed as a robust ground state for both iron-pnictides and iron-chalcogenides based on strong electron-electron correlation. However,  conclusions drawn in both studies are fundamentally unsatisfied.  For example, if the ground states in iron-based superconductors can approach different symmetries at different doping, it is hard to understand why the phase diagram of these materials is smooth against doping and  phase transitions between ground states with different pairing symmetries were not observed, and  a no-sign changed s-wave in iron-chalcogenides is not only inconsistent with observed neutron resonance but also generally unstable if it is driven by repulsive interaction. 

In a recent paper\cite{huoddparity},   one of us provided a complete symmetry classification  for  pairing symmetries in iron-based superconductors and showed that  spin-singlet pairing with odd parity can naturally take place  because of the intrinsic 2-Fe unit cell,  which was ignored in  previous theoretical studies based on  effective models with  1-Fe unit cell.  It was further argued that the pairing between electron pockets is most likely   $A_{1u}$($D_{2d}$) s-wave or $B_{2u}$ ($C_{4v}$) d-wave $\eta$-pairing.  The pairing state has both s-wave characters, such as no symmetry-protected nodes or nodal lines on  superconducting gap functions,  and d-wave characters, such as a sign change between top and bottom As/Se layers in real space. However, no microscopic mechanism has been proposed for odd parity pairing.

In this Letter, we provide a microscopic understanding to show that the superconducting state in iron-based superconductors is classified as an odd parity s-wave spin-singlet pairing state (OPS) in a single trilayer FeAs/Se, the building block of the materials.  This conclusion is only based on the consensus that As(Se) plays a critical role in driving superconductivity so that superconducting pairing is  determined locally by  d-p hybridization.   Under this consensus, in a low energy effective model with only  d-orbitals in an  iron square lattice,  we show that a normal pairing order defined between two iron sublattices is parity odd. Namely, it carries a sign change between top and bottom As/Se layers. This is because the effective nearest neighbor (NN) hopping  of $t_{2g}$ d-orbitals induced by d-p hybridization is generated through the anti-bonding state of  $p$-orbitals formed between top and bottom As/Se layers. As parity is a good quantum number,  a superconducting order parameter between two next NN (NNN) sites  must be an $\eta$-pairing.  Thus, in  the effective d-orbital models, a superconducting state that does not violate parity conservation must  include both normal and $\eta$ pairing. An OPS, classified as a $A_{1u}$($D_{2d}$) s-wave or $B_{2u}$ ($C_{4v}$) d-wave in a full lattice symmetry\cite{huoddparity},   is a combination of a s-wave normal pairing between two sublattices and  a s-wave $\eta$-pairing within the sublattices.  We show that the meanfield Hamiltonian of this state explains the dual s-wave and d-wave type characters  observed experimentally and unifies the description of  iron-pnictides\cite{Johnstonreview} and iron-chalcognides\cite{Guojg2010,Hesl2012,Liudefa2012,Tansy2013}.  The results  conclude that parity conservation was violated in  proposed superconducting states\cite{review-hirschfeld2011} in the past.  The confirmation of the  odd parity state will have a tremendous impact on understanding high-$T_c$ mechanism.

Before we start our analysis, we repeat  the definition of normal and $\eta$ pairing in\cite{huoddparity}. Let  $\vec k$ be momentum with respect to the  1-Fe unit cell in an iron square lattice.  The normal pairing refers to $(-\vec k, \vec k)$ pairing and the $\eta$-pairing refers to $(-\vec k,\vec k+Q)$ pairing where the momentum vector $Q=(\pi,\pi)$   is   a reciprocal lattice vector  in a 2-Fe unit cell. 

 \begin{figure}
\begin{center}
\includegraphics[width=1\linewidth]{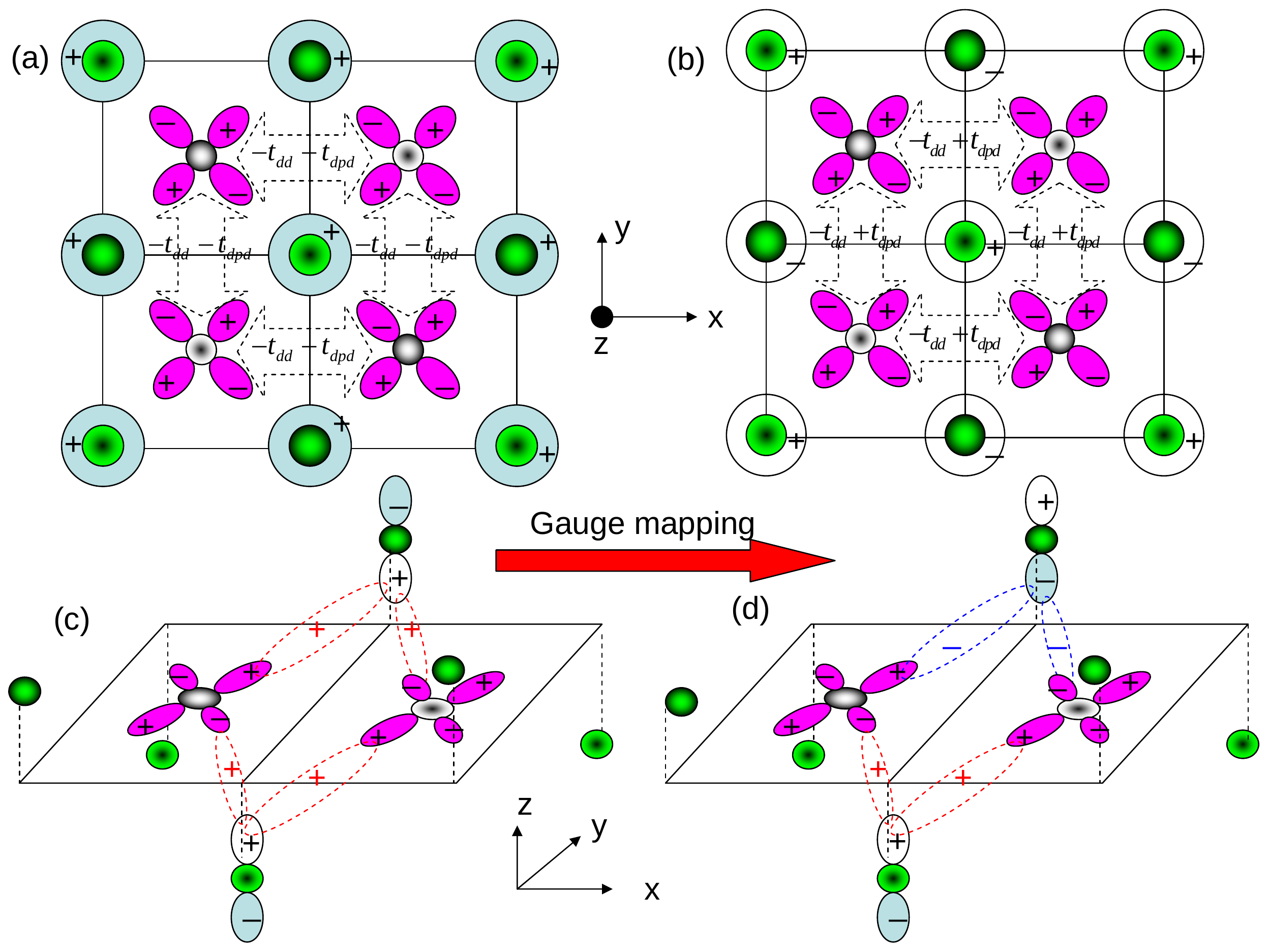}
\end{center}
\caption{(Color online) The nearest neighbor hopping parameters for intra-$%
d_{xy}$-orbital are shown in (a) and (b). $t_{dd}$ is the amplitude of the direct hopping of $%
d_{xy}$ orbital while $t_{dpd}$ is the amplitude of the indirect hopping through $p_{z}$
orbital of As/Se atom. The reason of sign change for $t_{dpd}$ between (a)
and (b) is that the $p_z$ orbitals in top layer and bottom layer  form
occupied bonding states in (a) and empty anti-bonding states in (b). The difference is illustrated by filled and empty $p_{z}$ orbitals in (a) and (b). (c) and (d) shows the local $p-d$ s-wave pairing
pattern  and the gauge transformation between them. }
\label{fig1}
\end{figure}
 {\it Effective Hamiltonian}
 We consider a general Hamiltonian in a single trilayer  $Fe-As(Se)$ structure coordinated by Fe and As(Se) atoms, 
\begin{eqnarray}
\hat H= \hat H_{dd} +\hat H_{dp} + \hat H_{pp}+\hat H_I
\end{eqnarray}
 where  $\hat H_{dd}$, $H_{dp}$ and $\hat H_{pp}$ describe the direct hopping between two d-orbitals,  the $d-p$ hybridization between Fe and As(Se)  and the direct hopping between two p-orbitals respectively.  $\hat H_I$ describes any  standard  interactions.   Here we do not need to specify the detailed parameters.  This Hamiltonian has a full symmetry defined by a non-symmorphic group  which can be specified equivalently as  $G=(\hat E, \hat I)\otimes C_{4v}$ or $G=(\hat E, \hat I) \otimes D_{2d}$ as shown in \cite{huoddparity}, where $\hat I$ is the space inversion operation defined at the center of  a NN Fe link, $D_{2d}$ is the point group at iron sites and $C_{4v}$ is the point group at the center of an iron square.

An effective Hamiltonian is obtained by integrating out  $p$-orbitals, which can be written as
\begin{eqnarray}
\hat H_{eff}= \hat H_{dd,eff} +\hat H_{I,eff} \label{eff}.
\end{eqnarray}
The effective band structure can be written as $\hat H_{dd,eff}= \hat H_{dd}+\hat H_{dpd}$, where $\hat H_{dpd}$ is  the effective hopping induced through d-p hybridization.  $\hat H_{dd,eff}$ has been obtained by many groups\cite{Kuroki2011,Miyake2009,Graser2009,Graser2010,Eschrig2009}. The major effective hopping terms  in $\hat H_{dpd}$ can be divided into two parts $\hat H_{dpd,NN} $,  which describes NN hopping and $\hat H_{dpd,NNN}$, which describes NNN hoppings in the iron square lattice. If one carefully checks the effective hopping parameters for  $t_{2g}$ orbitals in $\hat H_{dpd,NN}$,  one finds that they have opposite sign to what we normally expect  in a natural  gauge setting as shown in fig.\ref{fig1}(a,b), where $d_{xy}$ orbital is illustrated as an example. We see that  the hopping parameter $t_{dd}$ must be negative.  However   the effective hopping  parameter, $t_{dpd}$, is  positive and  even larger than $|t_{dd}|$ in \cite{Kuroki2011,Miyake2009,Graser2009,Graser2010,Eschrig2009}.  In a tetragonal lattice, $t_{dpd}$ can only be generated through  $d_{xy}-p_z$ hybridization. A positive value of $t_{dpd}$ suggests that  virtual hopping which generates $t_{dpd}$ must go through an unoccupied $p_z$ state. As shown in fig.\ref{fig1}(a,b),  a $d_{xy}$ equally couples to $p_z$  orbitals of top and bottom As atoms. A high energy $p_z$ state must be an anti-bonding $p_z$ state between NN As atoms. This analysis is held for all $t_{2g}$ orbitals which play the dominating role  in low energy physics.  It is also easy to check that the effective NNN hoppings between $t_{2g}$ orbitals are dominated through an occupied $p$ states, which is primarily a bonding state of $p$ -orbitals.  Therefore, the NN effective hoppings are generated through $d-p_a$ hybridization, where $p_a$ represents an anti-bonding p-orbital states and the NNN effective hoppings are generated through $d-p_b$ hybridization where $p_b$ is the bonding p-state.

\begin{figure}
\centerline{\includegraphics[width=1\linewidth]{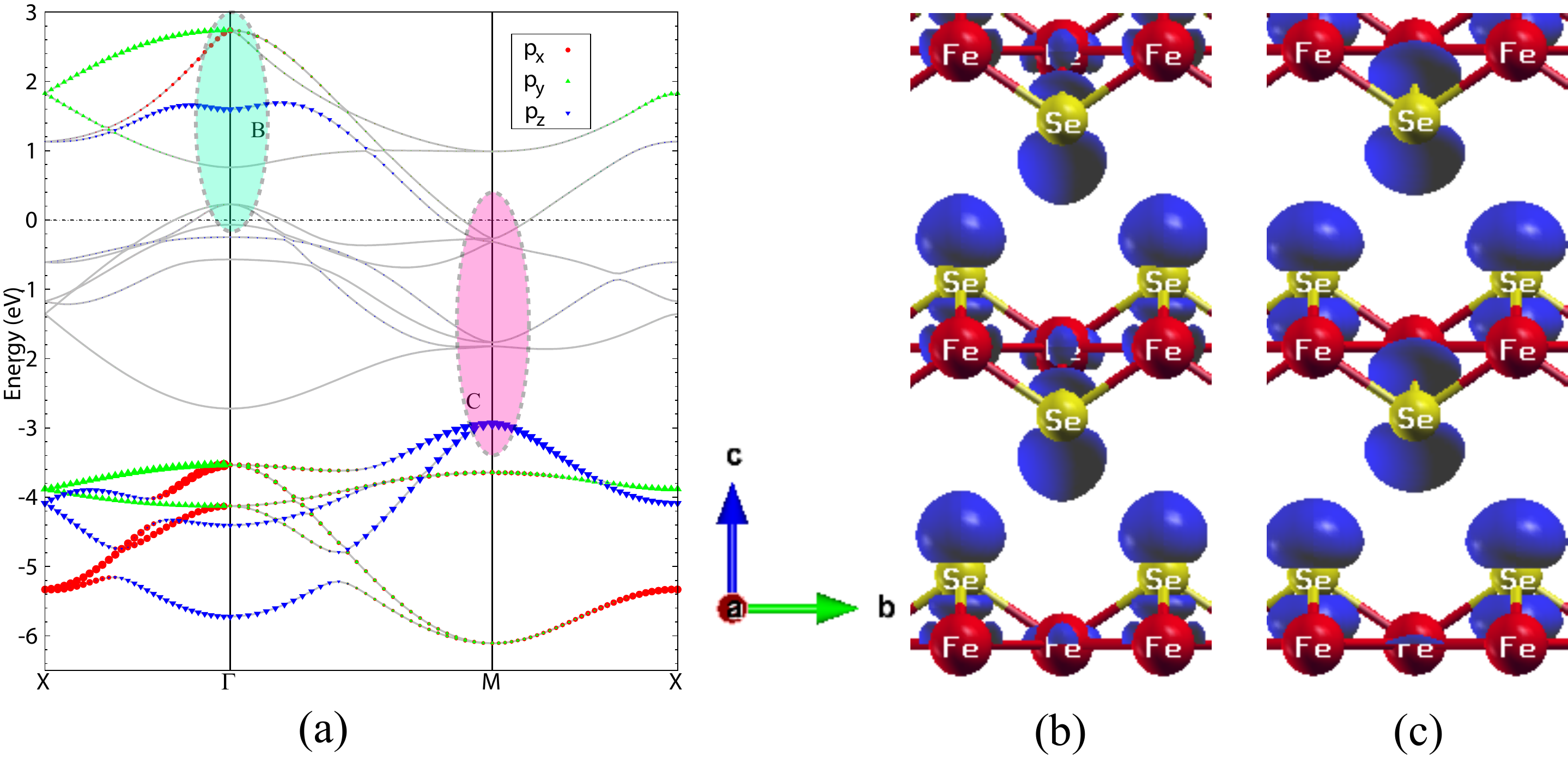}}
 \caption{ (a) The calculated band structure of  FeSe with the weight of p orbitals of Se. (b) The decomposed charge density  of  the band at $\Gamma$ marked by letter B (anti-bonding states). (c)The decomposed charge density  of  the band at $M$ marked by letter C (bonding states). 
 \label{fig2} }
\end{figure}

The above microscopic understanding is not surprising. In fact,  it is known in LDA calculations\cite{Miyake2009,Ma2008alu,singh2008a} that p-orbitals in As/Se are not fully occupied and there are significant overlappings between p-orbitals on bottom and top As/Se layers. Moreover, since $\hat H_{dpd,NN}$ and $\hat H_{dpd, NNN} $ primarily affect  hole pockets  around $\Gamma$ and electron pockets at $M$ separately, we can check the distribution of anti-bonding p states and bonding p states in  band structure to further confirm the analysis.   In fig.\ref{fig2}(a), we  plot the band structure of  FeSe and the distribution of $p$ orbitals. The $p_z$ orbitals of Se are mainly at +1.5 eV  at $\Gamma$ and -3 eV at $M$. By analyzing  the  bands at $\Gamma$ and $M$  as shown in Fig.\ref{fig2}(b) and (c), we confirm that  the   $p_z$ orbitals of Se at $\Gamma$ and $M$ belong to  anti-bonding and bonding states separately.

 {\it Hidden $Z_2$ symmetry characters in effective Hamiltonian:}
 % The effective Hamiltonian typically is viewed as  $C_{4v}$ symmetry which is not a symmetry of the original model.  However, 
Knowing the above hidden microscopic  origins  in a derivation of an effective Hamiltonian allows us to understand the symmetry characters of the effective Hamiltonian in the original lattice symmetry. As shown in\cite{huoddparity}, in  the original lattice symmetry,  $G$,  we can introduce a $Z_2$ classification specified  by $\hat \sigma_h$,  where $\hat \sigma_h$ is the reflection along z-axis. 
  
   The original Hamiltonian is invariant under $ (\sigma_h, \hat T)$, where  $\hat T$ is an in-plane translation by one Fe-Fe lattice. The $d-p_a$ hybridization is odd under $ \hat \sigma_h$ while the $d-p_b$ hybridization is even under $ \hat \sigma_h$. Thus,  the NN hopping $\hat H_{dpd,NN}$ and NNN  hopping $\hat H_{dpd,NNN}$ should be classified as odd and even under $\hat \sigma_h$ respectively. Namely,
   \begin{eqnarray}
  & & \hat \sigma_h  \hat H_{dpd,NN}  \hat \sigma_h=-1\nonumber \\
  & &   \hat \sigma_h  \hat H_{dpd,NNN}  \hat \sigma_h=1
  \label{symmetry}
   \end{eqnarray}
The above hidden symmetry property   is against the main  assumption taken in many weak coupling approaches, which assume that the essential physics is driven by the interplay between hole pockets at $\Gamma$ and electron pockets  at $M$ \cite{review-hirschfeld2011}.  As indicated in fig.\ref{fig2}(a),  the interplay between the hole and electron pockets must be minimal because of their distinct microscopic origins. 

{\it  Gauge transformation and parity of pairing order parameter:}
The symmetry difference in eq.\ref{symmetry} has fundamental impact on how to consider the parity  of a superconducting state if superconducting pairing is driven by local d-p hybridization.

  It has been shown that   in a system where  short range pairings in real space dominate, superconducting order parameters are momentum dependent and a gauge principle must be satisfied because the phases  of superconducting order parameters can be exchanged with those of the local hopping parameters\cite{bergkivelson2010,Hu2012s4} by gauge transformations.  As an example, a d-wave superconducting state in cuprates can be mapped to a s-wave superconducting state by a gauge mapping which changes the hopping terms from s-type symmetry to d-type symmetry\cite{Hu-s4review}. Therefore, only  the combined symmetry   of    hopping terms and  pairing orders associated to them  is a gauge-independent symmetry character to classify  states. This gauge principle does not exist in a conventional BCS-type superconductor in which the information of pairing in real space is irrelevant. 
\begin{figure}
\begin{center}
\includegraphics[width=1\linewidth]{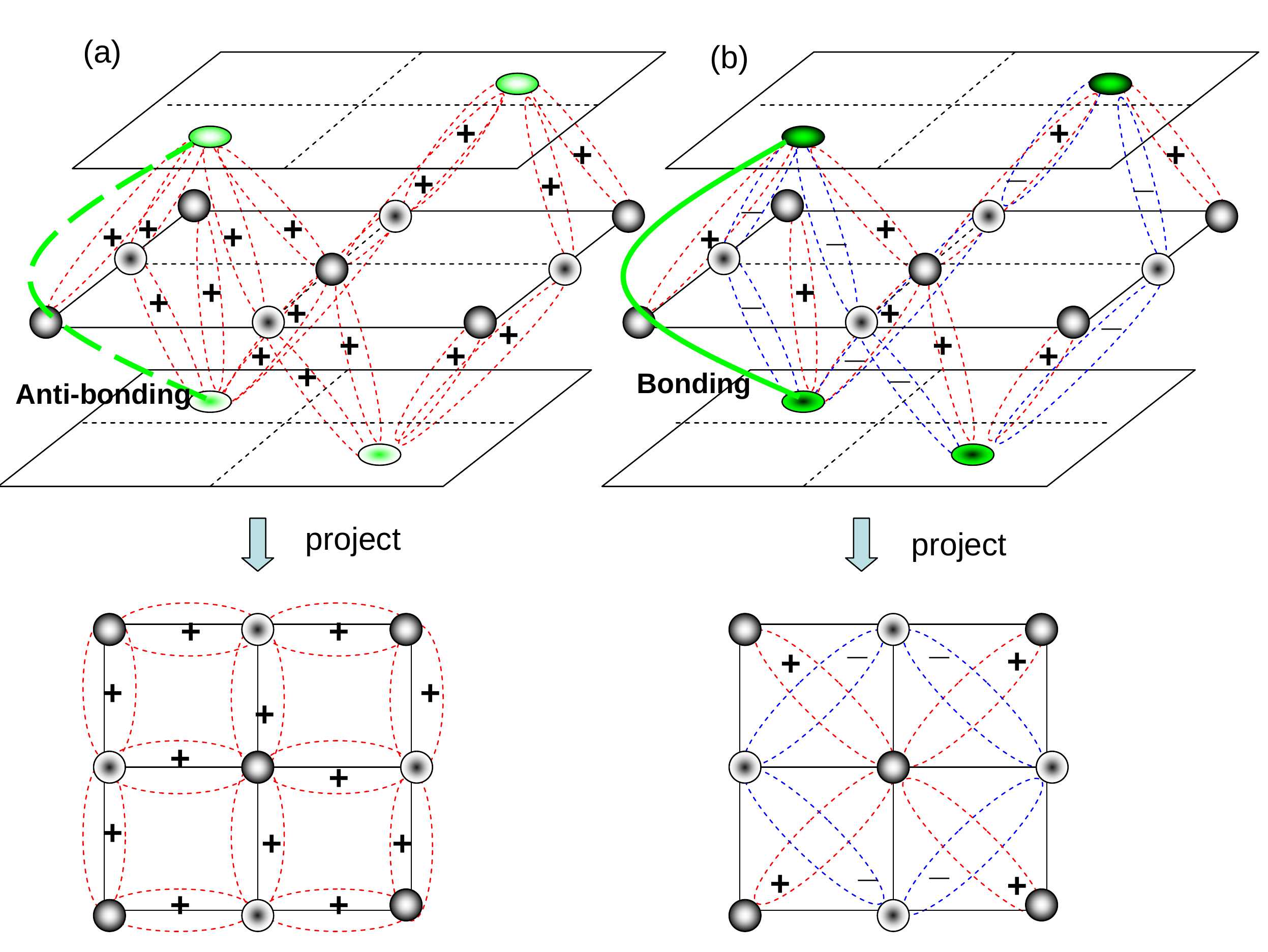}
\end{center}
\caption{(Color online) The NN and NNN $p-d$ local
pairing patterns with odd parity are shown in (a) and (b) in the natural gauge.  Note that p orbitals of As/Se in (a) form the anti-bonding
states while that in (b) form the bonding states. We distinguish the two
states with different filled green balls between (a) and (b). The $p-d$
pairings can be projected into effective $d-d$ pairings shown in the bottom
row. }
\label{fig3}
\end{figure}

Now we apply the gauge principle and  let $\hat \Delta_{NN}$ and $\hat \Delta_{NNN}$ be superconducting order operators associated with  $\hat H_{dpd,NN}$ and $\hat H_{dpd,NNN}$ respectively. In a superconducting state, we must have 
\begin{eqnarray}
[\hat \Delta_{NN}][\hat H_{dpd,NN}]=[\hat \Delta_{NNN}] [\hat H_{dpd,NNN}]
\label{gauge}
\end{eqnarray}
where $[\hat A]$ indicate the symmetry of  $\hat A$. Following eq.\ref{symmetry}, we  have $ [\hat \Delta_{NN}]=-[\hat \Delta_{NNN}] $ under $\hat \sigma_h$.  Therefore, based on the classification of pairing symmetries  in\cite{huoddparity},  we immediately conclude that the parity is odd   and the superconducting order $<\hat \Delta_{NNN}>$ must be an $\eta$-pairing  if $<\hat \Delta_{NN}>$ is a normal pairing.

The above analysis can be easily illustrated in real space. As shown in fig.\ref{fig3}, if superconducting pairing is driven by local d-p hybridization, the superconducting order  is  a pairing between d and p orbitals $\Delta_{dp}= <\hat d^+\hat p^+>$. A uniform $ <\hat d^+\hat p_a^+>$ is parity odd.  The NN pairing, $<\hat \Delta_{NN}>$ in the effective model, must originate from $ <\hat d^+\hat p_a^+>$ and thus is also parity odd.  The gauge principle  can be understood as shown in fig.\ref{fig1}(c,d).  If we can take a new gauge for Fermion operators of p-orbitals, $\hat p\rightarrow -\hat p $, in one of  the two As(Se) layers,   the anti-bonding operator $\hat p_a$ maps to the bonding operator $\hat p_b$. This gauge mapping  exactly transfers  the parity  between hopping terms and superconducting order parameters.

\begin{figure}
\begin{center}
\includegraphics[width=1\linewidth]{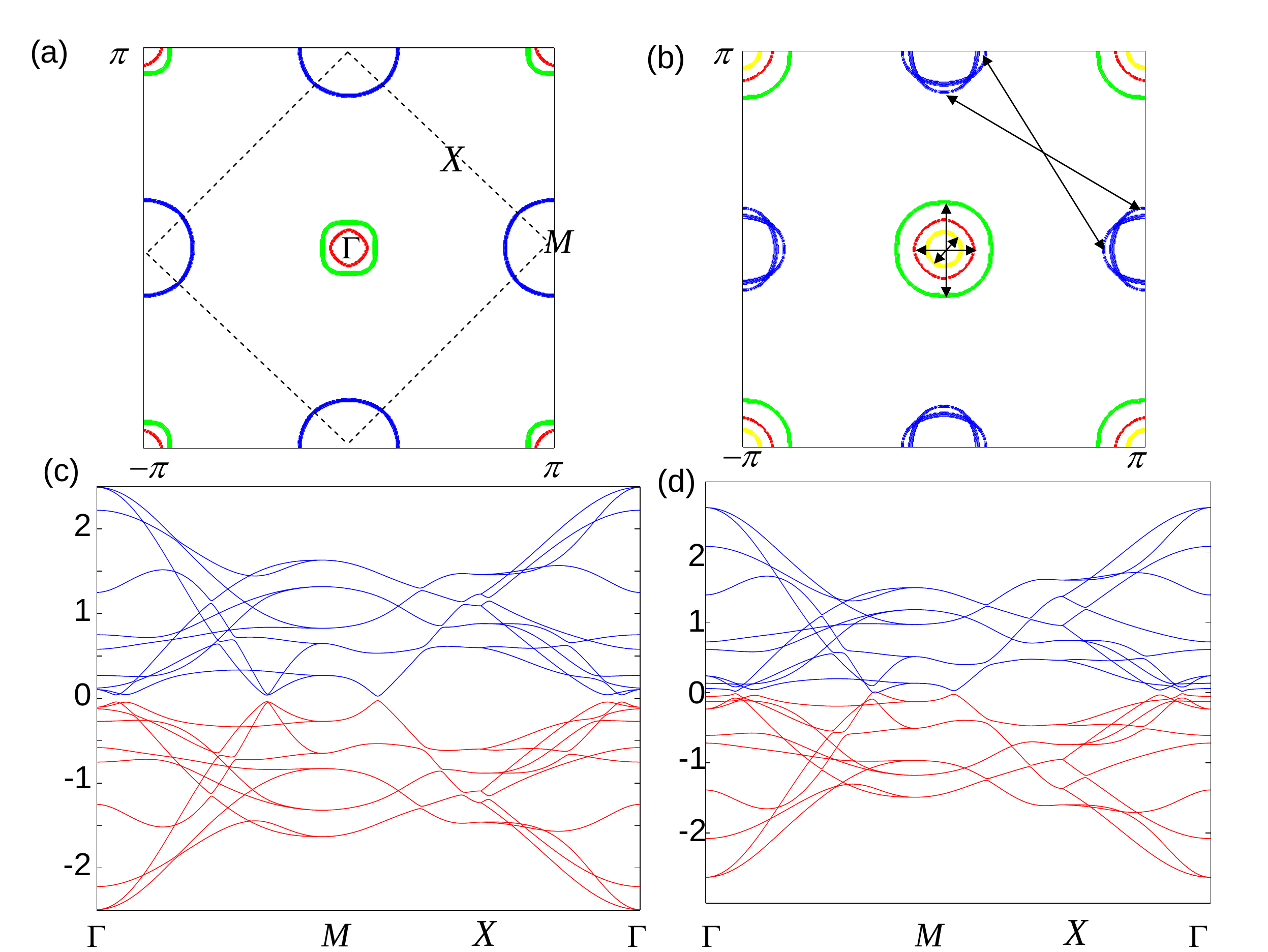}
\end{center}
\caption{(Color online) The Fermi surfaces of a five-orbital model in\cite{Graser2009}  are
shown in (a) and (b). The forms of hopping terms and hopping parameters can
be found  in\cite{Graser2009}. Here, we only add a chemical
potentials to tune the fermi level. We set $\protect\mu =0.1$ and $-0.04$ in
(a) and (b). The high-symmetry points are shown in (a) and the pairing
channels connecting the points on the fermi surface are denoted by the black
lines with arrows. The quasi-particle spectrum of the superconductive
states for (a) and (b) are shown in (c) and (d). We can find the (c) is full
gaped and (d) has nodes at the electron pockets. The superconductive order
parameters are chosen: $\Delta _{11,x}^{N}=\Delta _{11,y}^{N}=0.05;$ $\Delta
_{44}^{N}=0.05;$ $\Delta _{11}^{NN}=0.05;$ $\Delta _{12}^{NN}=0.05;\Delta
_{44}^{N}=-0.1;$ }
\label{fig4}
\end{figure}

 {\it Meanfield Hamiltonian for odd-parity s-wave state:}
 The above analysis can be generalized to all effective hoppings. The basic idea is to divide the iron square lattice into two sublattices. 
 In an odd parity s-wave state, the pairing between two sublattices must be normal pairing while the pairing within sublattices must be $\eta$-pairing.  Therefore,   a meanfield Hamiltonian to describe the odd parity s-wave state in 1-Fe unit cell can be generally written as
 \begin{eqnarray}
 H_{mf}= H_{dd,eff}+\sum_{\alpha,\beta, k}( \delta_{\alpha\beta, n} \hat \Delta_{\alpha\beta,n}(\vec k)+\delta_{\alpha\beta, \eta} \hat \Delta_{\alpha\beta,\eta}(\vec k)+h.c.)
\label{mf}
 \end{eqnarray}
where  $\hat \Delta_{\alpha\beta,n}= \hat d_{\alpha\uparrow}(\vec k)\hat d_{\beta\downarrow}(-\vec k)-d_{\alpha\downarrow}(\vec k)\hat d_{\beta\uparrow}(-\vec k) $ and $ \hat \Delta_{\alpha\beta,\eta}= \hat d_{\alpha\uparrow}(\vec k)\hat d_{\beta\downarrow}(-\vec k+Q)-d_{\alpha\downarrow}(\vec k)\hat d_{\beta\uparrow}(-\vec k+Q)$. In general, the normal   and $\eta$ pairing order parameters satisfy
\begin{eqnarray} & & \delta_{\alpha\beta, n} (\vec k)= -\delta_{\alpha\beta,n}(\vec k+Q) \\
& & \delta_{\alpha\beta, \eta} (\vec k)= \delta_{\alpha\beta,\eta}(\vec k+Q) .
\end{eqnarray}
These equations capture the sign change of superconducting order parameters in momentum space. The sign change here is required by symmetry. As the inter-orbital pairing can be ignored for s-wave pairing and the pairing is dominated by NN and NNN pairings, the important parameters are 
$\delta_{\alpha\alpha, n} \propto cosk_x+cosk_y$ and $\delta_{\alpha\alpha, \eta} \propto cosk_xcosk_y $. Thus, the superconducting gaps on hole pockets are mainly determined by  $\delta_{\alpha\alpha, n} $  and those on electron pockets are mainly determined by $\delta_{\alpha\alpha, \eta}$.   While detailed studies will be carried out in the future, the superconducting gaps obtained from eq.\ref{mf} can explain   experimental results observed in both iron-pnictides and iron-chalcogenides\cite{notes}. There is no symmetry protected node in this superconducting state. However, accidental nodes can easily take place.  In fig.\ref{fig4}, we plot numerical results for two cases.  Parameters are specified in the caption of the figure.      The superconducting gap  in the first case is a  full gap while it has gapless nodes on electron pockets in the second case.

{\it  Impact to High-$ T_c$ Mechanism}
In ref.\cite{huoddparity}, the odd parity s-wave state was conjectured  based on intriguing experimental facts.  With microscopic mechanism proposed here,  we can address important impact on high $T_c$ mechanism  for  iron-based superconductors and other high $T_c$ superconductors  if the state is confirmed.

First,   the microscopic mechanism revealed here   fundamentally  differs from those proposed in weak coupling approaches which only emphasize Fermi surfaces. Fermi surfaces   are only determined by energy dispersion. It provides no information about   underlining microscopic processes which are  local and  bound with high energy physics.  In correlated electron systems, these processes essentially  determine many important properties.

Second, our study provides fundamental reasons why we failed to recognize the odd parity symmetry in the past.  The  symmetry principle and gauge principle were not properly handled. In the past,  the effective Hamiltonian was viewed  in the symmetry group $C_{4v}$ at iron sites rather than the original lattice symmetry $G$.  We can see that if $\hat \sigma_h$ could  be set to one, $G$ is reduced to $C_{4v}$. However,  due to the  anti-bonding $p$ orbital states,  the effective Hamiltonian does not represent correct symmetry of the original lattice in a natural gauge setting.  The correct physics can only  be understood after the hidden gauge is revealed. For  an order parameter which is momentum dependent,  this gauge information is critical.    The gauge principle becomes very important  for us to search new physics in other complex electron  systems.

Finally, if the odd parity state is confirmed,  the fundamental objects  in   superconducting  states of high $T_c$ materials must be the  tightly binding Cooper pairs between d and p orbitals.  In this view, the odd parity s-wave state closely resembles the d-wave state in a Cu-O plane of cuprates. We expect an  identical mechanism to select sign changed superconducting orders in both materials.

In summary,  we provide a microscopic mechanism to support the formation of an odd parity s-wave superconducting state in iron-based superconductors.  We demonstrate that  in an effective model based on d-orbitals,   both normal pairing and $\eta$ pairing must be included if the superconducting state conserves parity.    Superconducting states studied in the past violate parity conservation.

{\it Acknowledges:}   JP  acknowledges  H. Ding and D.L. Feng for useful discussion.  The work is supported  by the Ministry of Science and Technology of China 973 program(2012CB821400) and NSFC-1190024.

%\emph{ Acknowledgments:}
%The author acknowledges  H. Ding and P. Coleman for extremely stimulated discussion,   NN Hao for useful discussion and help on figures, and   D. Scalapino, T. Xiang, X. Dai,  and D.L. Feng for useful discussion.  The work is supported  by the Ministry of Science and Technology of China 973 program(2012CB821400) and NSFC-1190024.

%\bibliography{FeAspc,myownpaper}

\begin{thebibliography}{22}
\bibitem{notes} The analysis has been provided in ref.8.
\end{thebibliography}

\end{document}